\documentclass[twocolumn]{aastex631}

\def\coordra{11:38:10.91~}
\def\coorddec{$-$51:39:49.15~}
\def\fullcoords{\coordra \coorddec}
\def\shortcoords{SMSS J1138$-$5139}
\def\msol{{\rm M_\odot}}
\def\rsol{{\rm R_\odot}}

\def\ptessmin{13.84~{\rm min}}
\def\ptessminb{27.6797\pm0.0009}

\def\prvb{27.69\pm0.03~{\rm min}}
\def\prvbb{27.69\pm0.03}

\def\loggb{6.31\pm0.18}
\def\loggbldc{6.31}

\def\vrotb{255\pm50~{\rm km~s^{-1}}}

\def\metalb{-0.54\pm0.2}

\def\mb{0.40^{+0.22}_{-0.15}~\msol}

\def\minmass{1.21^{+0.22}_{-0.15}~\msol}

\def\tmerge{5.7\pm0.3~{\rm Myr}}

\def\kb{687\pm13~{\rm km~s^{-1}}}
\def\kbb{687\pm13}
\def\gammab{+59\pm6~{\rm km~s^{-1}}}
\def\gammabb{+59\pm6}

\def\rb{0.073^{+0.007}_{-0.006}~\rsol}

\def\teffb{9650\pm300~{\rm K}}
\def\teffbb{9650\pm300}
\def\teffbldc{9650~{\rm K}}

\def\parallax{1.83\pm0.06~{\rm mas}}

\def\qlcurve{0.24\pm0.01}

\def\mblcurve{0.24\pm0.01~\msol}
\def\malcurve{0.99\pm0.01~\msol}

\def\incb{88.7~{\rm deg}}
\def\mblcurveb{0.24~\msol}
\def\malcurveb{0.99~\msol}

\def\incc{88.7\pm0.1}
\def\mblcurvec{0.24\pm0.01}
\def\malcurvec{0.99\pm0.01}
\def\rblcurvec{0.0859\pm0.0005}

\def\lisasnr{7-10}

\usepackage{hyperref}

\shorttitle{SMSS J1138$-$5139}
\shortauthors{Kosakowski et al.}

\graphicspath{{./}}

\begin{document}

\title{A New LISA-Detectable Type Ia Supernova Progenitor in the Southern Sky: SMSS J1138$-$5139}

\author[0000-0002-9878-1647]{Alekzander Kosakowski}
\affiliation{Department of Physics and Astronomy, Texas Tech University, 2500 Broadway, Lubbock, TX, 79409, USA}
\affiliation{Department of Physics and Astronomy, University of North Carolina, Chapel Hill, NC, 27514, USA}

\author[0000-0001-5400-2368]{Matti Dorsch}
\affiliation{Institut für Physik und Astronomie, Universität Potsdam, Karl-Liebknecht-Str. 24/25, 14476 Potsdam, Germany}

\author[0000-0002-4462-2341]{Warren R. Brown}
\affiliation{Center for Astrophysics, Smithsonian Astrophysical Observatory
60 Garden St.,
Cambridge, MA, 012138, USA}

\author[0000-0002-6540-1484]{Thomas Kupfer}
\affiliation{Department of Physics and Astronomy, Texas Tech University, 2500 Broadway, Lubbock, TX, 79409, USA}
\affiliation{Hamburger Sternwarte, University of Hamburg, Gojenbergsweg 112, 21029 Hamburg, Germany}

\author{Fatma Ben Daya}
\affiliation{Hamburger Sternwarte, University of Hamburg, Gojenbergsweg 112, 21029 Hamburg, Germany}

\author[0000-0001-6098-2235]{Mukremin Kilic}
\affiliation{Homer L. Dodge Department of Physics and Astronomy University of Oklahoma
1667 K Street NW, Suite 800 
Norman, OK 73072, USA}

\begin{abstract}
We present the discovery and analysis of a nearby eclipsing ultra-compact accreting binary at coordinates \fullcoords\ (\shortcoords), the first well-constrained LISA-detectable Type Ia supernova progenitor. Our time series optical spectroscopy identifies its orbital period through radial velocity monitoring at $P_{\rm orb, RV}=\prvb$; twice the photometric period seen in 2-minute cadence data from TESS Sector 37. We model our optical spectroscopy together with new simultaneous multi-band time series photometry from Gemini to place constraints on the binary parameters. Our light curve modeling finds that \shortcoords\ contains an $M_2=\mblcurveb$ pre-white dwarf donor with a massive $M_1=\malcurveb$ white dwarf accretor at orbital inclination $i=\incb$. Based on our photometrically-derived system parameters, we expect that gravitational wave radiation will drive \shortcoords\ to a merger within $\tau=\tmerge$ and result in a Type Ia supernova. Even without a direct merger event, the component masses of \shortcoords\ and active hydrogen accretion suggest that eventual helium accretion will likely also trigger a Type Ia supernova explosion through the dynamically-driven double-degenerate double-detonation (D6) channel. We expect LISA to detect the gravitational wave emission from \shortcoords\ with signal-to-noise $\lisasnr$ after a 48-month mission.

\end{abstract}


\section{Introduction} \label{sec:intro}
\begin{figure*}
    \centering
    \includegraphics[width=\textwidth]{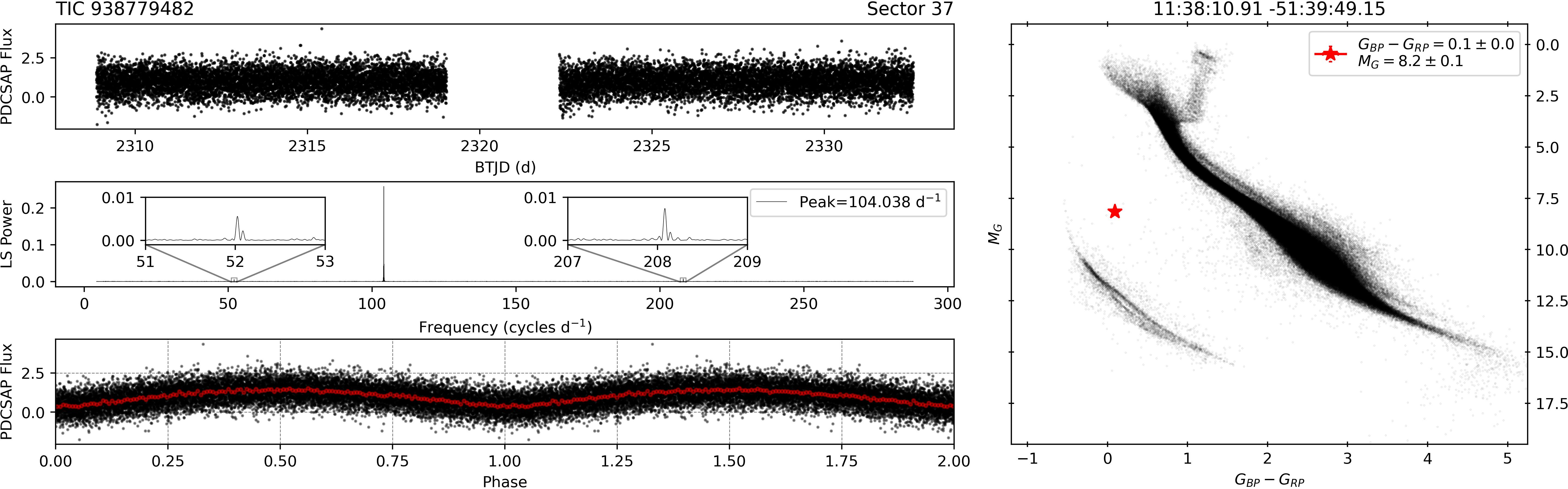}
    \caption{Left: Calibrated PDCSAP Flux 2-minute cadence TESS sector 37 light curve of \shortcoords\ (top), its Lomb-Scargle power spectrum (middle), and phase-folded 2-minute cadence TESS sector 37 light curve (bottom). We include zoomed inset plots showing significant peaks at half and twice the dominant peak. Right: The location of \shortcoords\ on the Gaia DR3 color-magnitude diagram.} 
    \label{fig:tess_als}
\end{figure*}

Ultra-compact binaries are a subclass of evolved post-common envelope binaries with orbital periods $P_{\rm orb}<1~{\rm h}$ and which may experience rapid orbital decay through a combination of gravitational wave emission, mass transfer, and tidal interactions. The formation, evolution, and potential merger of ultra-compact binaries has important astrophysical implications for common envelope evolution \citep{scherbak2023}, tidal interactions \citep{piro2011}, gravitational wave astronomy \citep{amaro2023}, the formation of stably accreting binaries \citep[i.e. AM CVn][]{nelemans2001, kilic2014, kilic2016}, and stellar mergers leading to the formation of various exotic sources, including He-rich stars \citep[i.e. R Coronae Borealis;][]{webbink1984,zhang2014}, massive single white dwarfs \citep{cheng2020,jewett2024}, and even Type Ia supernovae \citep{webbink1984, shen2024}.

Spectroscopic surveys targeting white dwarf binaries, most notably the Extremely Low Mass (ELM) white dwarf survey \citep[][and references therein]{brown2010, brown2022, kosakowski2023}, have identified ultra-compact binaries through time series radial velocity monitoring of ELM white dwarfs, including (up to orbital period of $30~{\rm min}$) J0651$+$2844 with $P_{\rm orb}=12~{\rm min}$ \citep{brown2011}, J0935$+$4411 with $P_{\rm orb}=20~{\rm min}$ \citep{kilic2014b}, J2322$+$0509 with $P_{\rm orb}=20~{\rm min}$ \citep{brown2020},  J1239$-$2041 with $P_{\rm orb}=22~{\rm min}$ \citep{brown2022}, and J0634$+$3803 and J0338$-$8139 with orbital periods $P_{\rm orb}=26~{\rm min}$ and $P_{\rm orb}=30~{\rm min}$, respectively \citep{kilic2021}.

More recently, time domain sky surveys, such as the all-sky Transiting Exoplanet Survey Satellite \citep[TESS;][]{ricker2015} and the northern sky Zwicky Transient Facility \citep[ZTF;][]{bellm2019,graham2019,masci2019}, have enabled an accelerated discovery rate of ultra-compact binaries through detections of periodic photometric variability caused by eclipses and tidal-deformation. \citet{burdge2020} have identified 15 such binaries in a targeted search for ultra-compact binaries in ZTF data, demonstrating the effectiveness of these time domain surveys in identifying ultra-compact binaries.

Archival data from time domain surveys such as the Palomar Transient Factory \citep[PTF;][]{law2009}, the All-Sky Automated Survey for Supernovae \citep[ASAS-SN;][]{kochanek2017}, and the Asteroid Terrestrial-impact Last Alert System \citep[ATLAS;][]{tonry2018,heinze2018} have been used to augment newer data sets to measure the effects of general relativity on the orbital decay of ultra-compact binaries through precise measurements of eclipse timing variations over many-year baselines \citep[see][]{hermes2012,burdge2019a,burdge2019b,burdge2020,burdge2023,chakraborty2024}.

Upcoming space-based gravitational wave detectors, such as the Laser Interferometer Space Antenna \citep[LISA;][]{amaro-seoane2017}, TianQin \citep{luo2016}, and Taiji \citep{ruan2018}, will be sensitive gravitational wave emission at mHz frequencies from merging ultra-compact binaries. Detailed simulations of Galactic binary populations find that LISA is expected to individually resolve gravitational wave signal from $\mathcal{O}(10^4)$ Galactic binaries \citep{korol2017, korol2022}, while upwards of $\mathcal{O}(10^2)$ of these are also expected to be identified and characterized through their electromagnetic radiation as ``multi-messenger" sources \citep{breivik2018,korol2019,li2020}.

To date, approximately 43 LISA-detectable binaries have been characterized through their electromagnetic radiation \cite[see][and references therein]{finch2023,kupfer2024,chakraborty2024}. Due to the relative abundance of northern sky follow-up facilities and sky surveys, most of these known binaries are located in the northern sky. However, future southern sky time domain surveys such as BlackGEM \citep{bloemen2015, groot2024}, the La Silla Schmidt Southern Survey\footnote{\url{https://sites.northwestern.edu/ls4/}} (LS4), and the Vera Rubin Observatory Legacy Survey of Space and Time \cite[LSST;][]{ivezic2019} will create a similar catalog of transients and variables in the southern sky.

\begin{figure*}[t!]
    \centering
    \includegraphics[width=\textwidth]{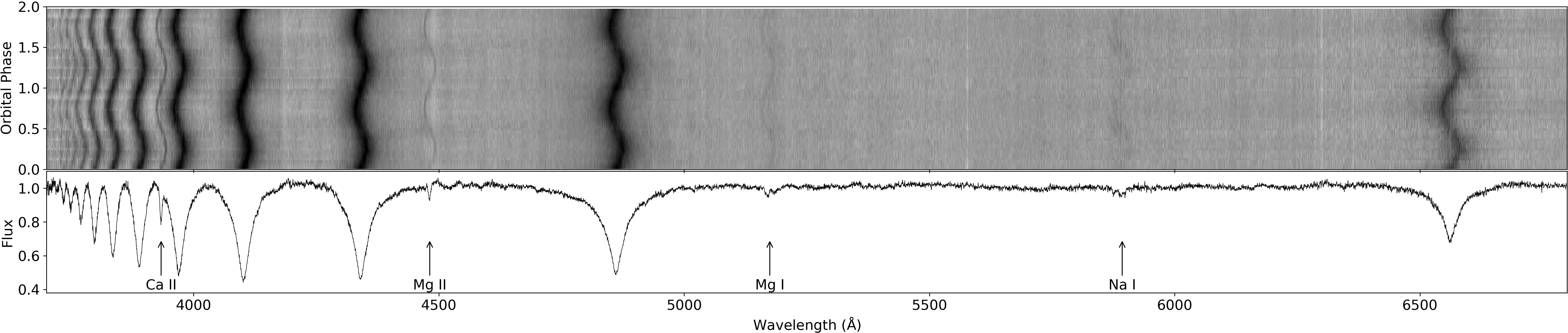}
    \caption{Top: Trailed MagE optical spectrum (orbital phase vs wavelength vs normalized flux) of \shortcoords\ covering wavelength range $3700-6800~{\rm \AA}$. Absorption components from Ca II ($3933~{\rm \AA}$), Mg II ($4481~{\rm \AA}$), the Mg I triplet ($5167,5173,5184~{\rm \AA}$), and the Na I doublet ($5890,5896~{\rm \AA}$) can be seen moving in sync with the orbital motion of the bright hydrogen-rich donor star. Bottom: Continuum-normalized co-added zero-velocity spectrum of \shortcoords. 
    The broad stationary features around $4500-4700~{\rm \AA}$ are thought to be an artifact of the Xe-flash flat field spectrum used to calibrate the MagE blue data. 
    } 
    \label{fig:trailed_spec_3700_7000}
\end{figure*}
\begin{figure}
    \centering
    \includegraphics[width=\columnwidth]{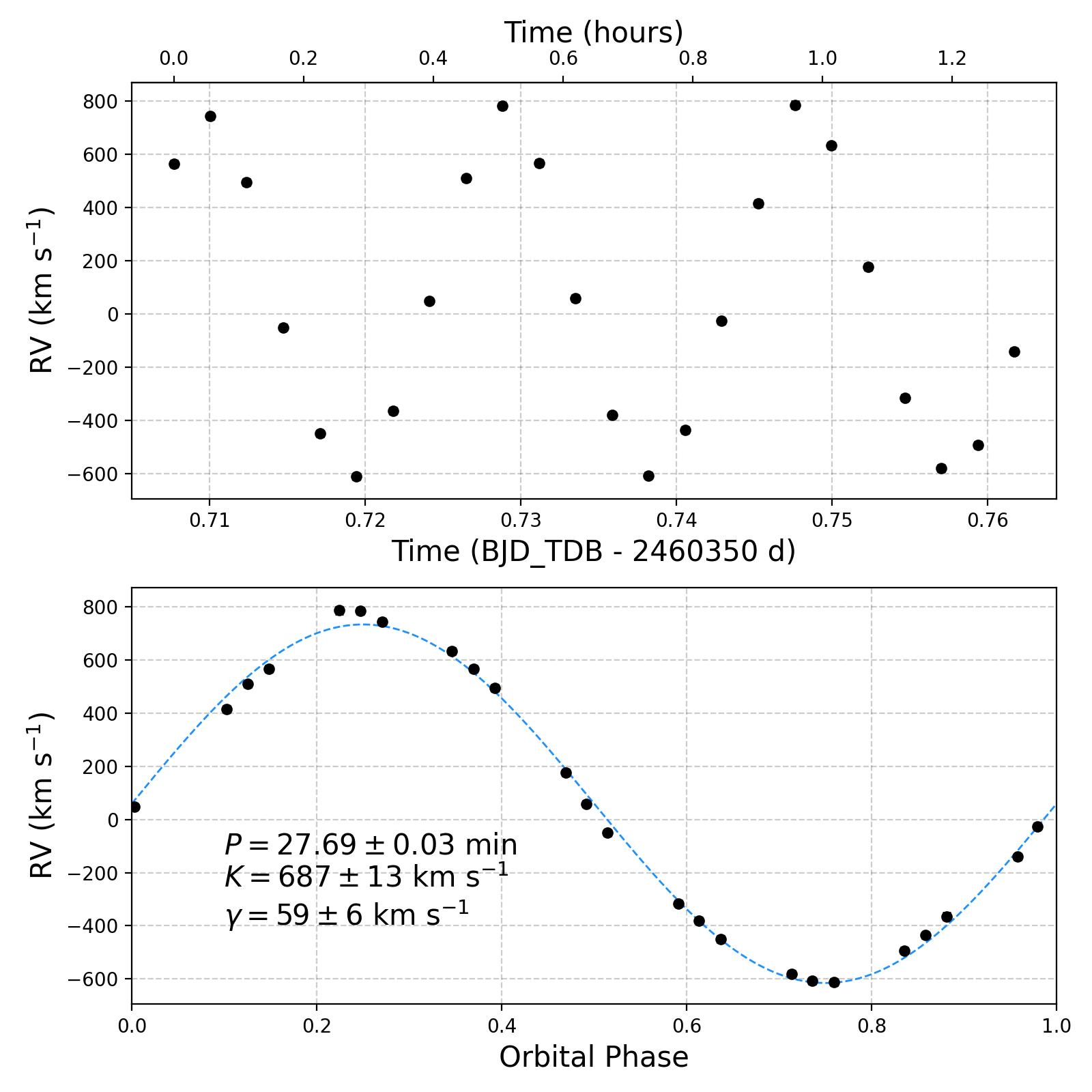}
    \caption{Top: Individual radial velocity measurements of \shortcoords\ taken using the MagE spectrograph on the 6.5-meter Magellan Baade telescope. Given the scale of velocity semi-amplitude, the errorbars appear smaller than the individual data points.  Bottom: Best-fitting integrated sine function circular orbit fit to the measured radial velocities of \shortcoords.}
    \label{fig:rv_plot}
\end{figure}

Here we present our analysis of a new bright, nearby, eclipsing, southern sky, ultra-compact, accreting binary based on new time series spectroscopic and photometric observations. Our spectroscopic and radial velocity analysis is presented in Section \ref{sec:spec}. Our analysis of the archival spectral energy distribution and new light curve modeling is presented in Section \ref{sec:phot}. Finally, we discuss the expected gravitational wave signal and eventual fate of \shortcoords\ in Section \ref{sec:disc}.

\section{Target Identification}
We identified \shortcoords\ as a bright (Gaia $G=16.9~{\rm mag}$) and nearby (Gaia DR3 $\varpi=\parallax$) southern sky extremely low mass white dwarf candidate based on archival color and astrometry selections using the SkyMapper DR2 \citep{onken2019} and Gaia DR3 \citep{gaia_dr3} data archives. \shortcoords\ is flagged with \textsc{phot\_variable\_flag=variable} within Gaia DR3, which is confirmed through its periodic photometric variability seen at $P_{\rm TESS}=\ptessmin$ in the 2-minute cadence data from the TESS Sector 37 data archive. 
We present the TESS Sector 37 2-minute cadence light curve of \shortcoords\ (TIC 938779482) in Figure \ref{fig:tess_als}.

\section{Spectroscopic Analysis}\label{sec:spec}
We obtained 24 back-to-back spectra covering approximately 80 minutes of observations with exposure times of ${\rm EXPTIME}=180~{\rm s}$ on UT 2024 February 10 (airmass $1.19-1.10$) using the Magellan Echellette \citep[MagE;][]{marshall2008} spectrograph on the 6.5m Magellan Baade telescope at the Las Campanas Observatory in Chile. Our observing configuration used the $0.85~{\rm arcsec}$ slit, providing wavelength coverage $3700-9300~{\rm \AA}$ at resolving power $R=4800$. We obtained a ThAr calibration arc lamp immediately after our observing sequence to ensure our spectra have an accurate wavelength solution. We note that our MagE spectra show broad stationary features around $4500-4700~{\rm \AA}$, which we believe to be artifacts from a Xe-flash flat field spectrum used for calibrating the MagE blue data.

Our spectroscopy reveals that the binary's visible component is dominated by broad H absorption lines and shows clear absorption components from Na, Mg, and Ca, which move in-sync with the H lines of the visible star, suggesting that the visible star is a cool DAZ spectral type. We display our phase-resolved trailed spectra and the zero-velocity co-added continuum-normalized spectrum for \shortcoords\ between $3700-6800~{\rm \AA}$ in Figure \ref{fig:trailed_spec_3700_7000}.

Our spectroscopy shows evidence for faint emission peaks in the cores of the hydrogen Balmer lines, most prominent around orbital phase $\phi\approx0.50-0.75$ (starting when the visible star is furthest away in its orbit; see Figure \ref{fig:hrs:full2} in the Appendix), suggesting the presence of a cool accretion disc eclipsing a bright donor star.

\subsection{Radial Velocity Analysis}

We estimated the observed radial velocity for each of our 24 spectra using a cross-correlation technique with the cross-correlation package \textsc{rvsao.xcsao} \citep{kurtz1998} within \textsc{iraf} \citep{tody1986}. We created a template spectrum for \shortcoords\ by shifting each of our 24 individual spectra to zero velocity and co-adding them into a single high-quality spectrum. We then cross-correlated the hydrogen Balmer lines in each of the individual spectra against this zero-velocity template spectrum to estimate their radial velocities. Finally, we applied the barycentric correction to our measured velocities and added in quadrature a systematic error of $\sigma_{v,{\rm sys}}=10~{\rm km~s^{-1}}$ to our measured uncertainties, based on similar velocity analyses of compact binaries.

\begin{figure}
    \centering
    \includegraphics[width=\columnwidth]{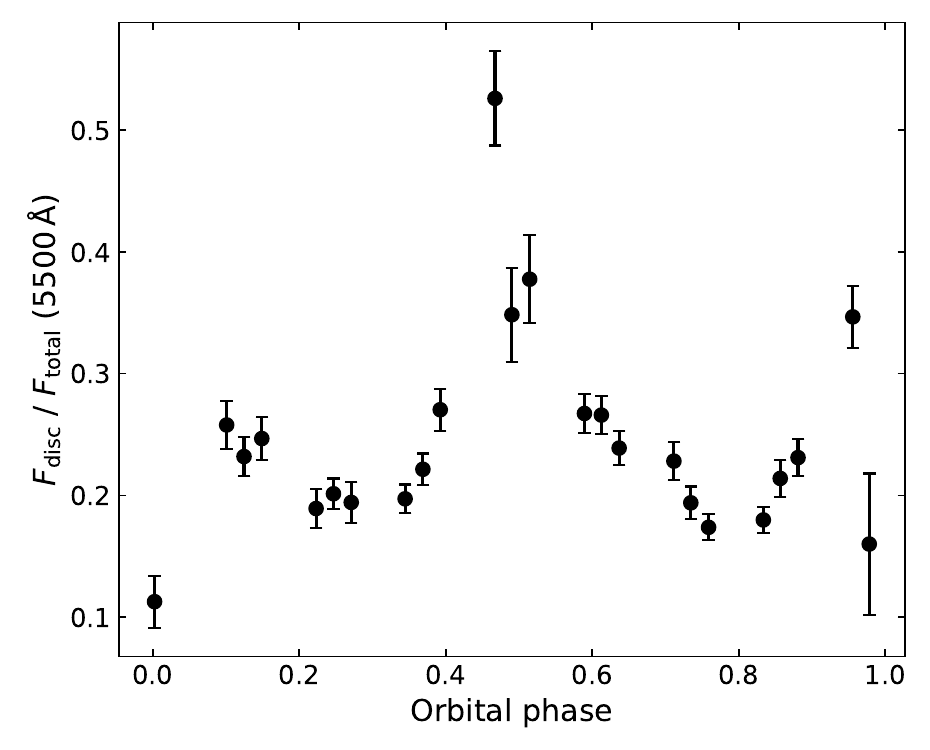}
    \caption{The resulting flux contribution of the disc component at 5500\,\AA\ relative to the total flux. The estimated disc contribution is strongest at $\phi=0.47$ due to an increase in its best-fit temperature to $T_\mathrm{eff}=7750\pm50\,\mathrm{K}$, which may be related to the accretion hot spot.}
    
    \label{fig:surface_ratio}
\end{figure}

\begin{figure*}
    \centering
    \includegraphics[width=\textwidth]{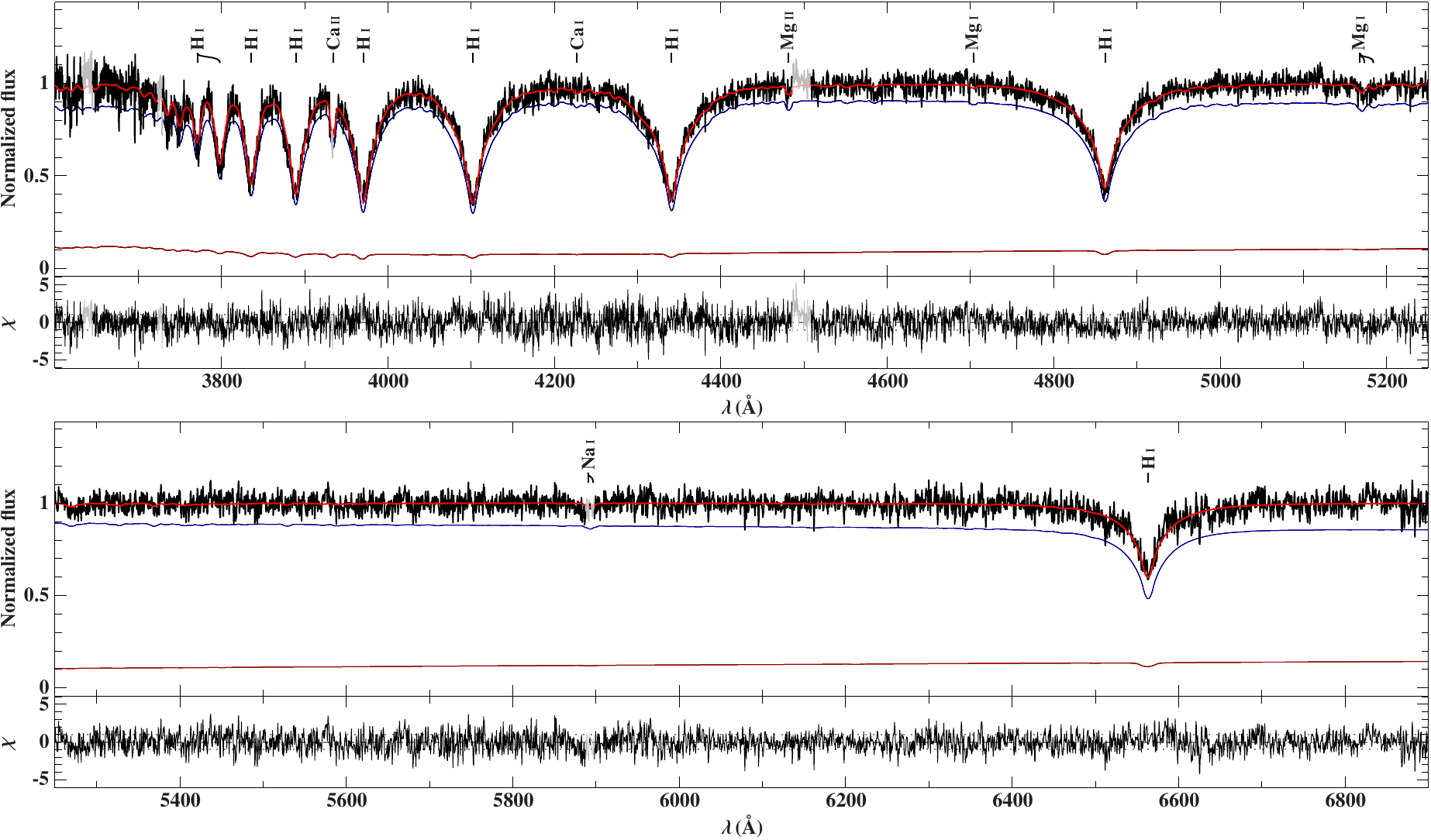}
    \caption{Best-fitting model spectrum to an individual MagE spectrograph optical spectrum for \shortcoords\ at orbital phase $\phi=0$. The individual relative contributions from the donor (blue) and accretor / disc (dark) are shown below the observed spectrum (black) and the best-fit combined model (red). The bottom panels show uncertainty-weighted residuals $\chi$. 
    }
    \label{fig:spec_fit}
    \vspace{4pt}
\end{figure*}

We fit a circular orbit to our radial velocity measurements using an integrated sine function to account for orbital smearing in our data caused by our long exposures relative to the short orbital period, with each exposure covering approximately $11\%$ of the orbit. We find best-fitting orbital parameters: velocity semi-amplitude $K_2=\kb$, systemic velocity $\gamma=\gammab$, and orbital period $P_{\rm orb, RV}=\prvb$; twice the photometric period seen in TESS. Our radial velocity measurements and the best-fitting circular orbit is presented in Figure \ref{fig:rv_plot}.

\subsection{Atmospheric Parameters}
\label{sect:spec}

We used  $\chi^2$ minimization to fit our 24 individual MagE spectra against a grid of two-component ATLAS12 \citep{Kurucz1996} + SYNTHE \citep{Kurucz1993} pure 1D-LTE models which include metal line opacities. Level dissolution, an important effect for high-gravity stars, 
is considered using the implementation of \cite{Irrgang2018}. For simplicity, we approximate the disc component with a rapidly rotating stellar spectrum. This is possible because the spectrum is dominated by the donor at almost all phases. As shown in Figure \ref{fig:surface_ratio}, the disc only contributes about $22\pm3$\,\% of the total flux at 5500\,\AA\ when the donor is not eclipsed, despite being similar in apparent size.

All 24 available MagE spectra were fit simultaneously, forcing a common metallicity while allowing variable radial and rotational velocity, surface gravity, effective temperature and surface ratio. We find the following parameters for the visible donor star: surface gravity $\log{g_{\rm donor}}=\loggb$, effective temperature $T_{\rm eff,donor}=\teffb$, projected rotational velocity $v_{\rm rot,donor}\sin i=\vrotb$ and metallicity $\mathrm{[Fe/H]}=\metalb$. However, our rotational velocity and surface gravity may be overestimated due to smearing in the spectra.

The disc component is best reproduced at $T_{\rm eff,disc}=6810\pm 210\ \mathrm{K}$ and features a surface ratio relative to the donor of $A_\mathrm{disc}/A_\mathrm{donor} = 1.00 \pm 0.24$. 
These parameters represent average best-fit values obtained from out-of-eclipse spectra. 
An exception is the donor's surface gravity ($\log{g_{\rm donor}}$) which was estimated from the $\phi=0$ spectrum (at primary conjunction). This spectrum is least affected by the disc contribution due to the eclipse of the disc by the donor star. 
Estimated systematic uncertainties of 0.15 in $\log g$, 2.5\,\% in $T_\mathrm{eff}$, and 0.15 in [Fe/H] were added in quadrature to the statistical uncertainties. 
These substantial uncertainties are motivated by the relative simplicity of our models.
Most importantly, the non-spherical shape of the donor star is not accounted for and the disc contribution may introduce biases that are difficult to quantify. 

We present our best-fitting model spectrum overplotted onto our observed spectrum at orbital phase $\phi=0$ in Figure \ref{fig:spec_fit}, where the individual contribution from the donor/accretor is displayed below the fit as a blue/red line. As demonstrated in Figure \ref{fig:surface_ratio}, the relative contribution from the accretion disc to the total system flux varies as a function of orbital phase. For completeness, we present best-fitting model spectra at additional orbital phases in Figures \ref{fig:hrs:full}, \ref{fig:hrs:full2}, and \ref{fig:hrs:full3} in the Appendix.

\begin{figure}
    \centering
    \includegraphics[width=\columnwidth]{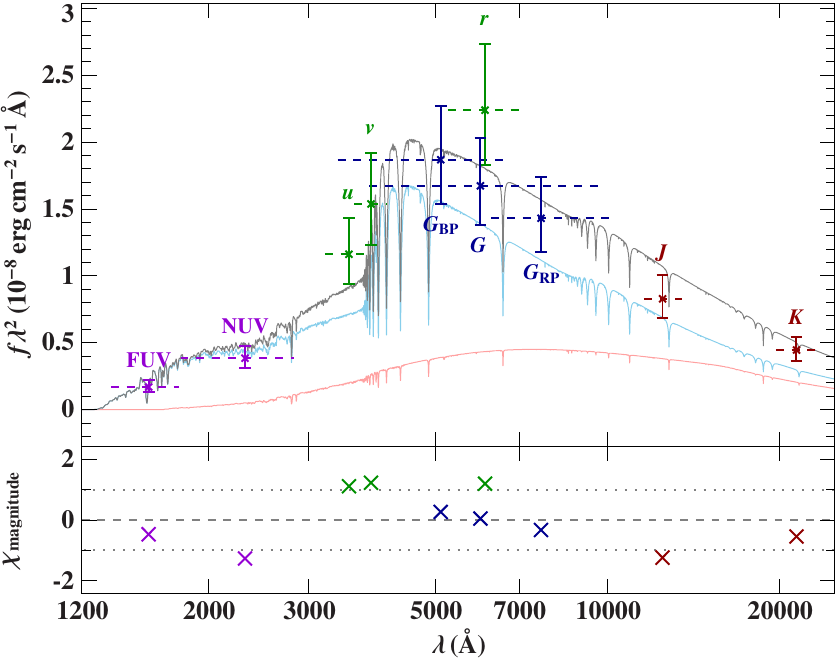}
    \caption{
Photometric fit for \shortcoords. Filter-averaged fluxes are shown as coloured data points and filter widths are indicated by dashed horizontal lines.
The grey line visualises the combined model spectrum while individual contributions are shown in blue (donor) and red (disc). 
The following colour codes are used to identify the photometric observations: 
GALEX \cite[violet,][]{Bianchi2017},
SkyMapper \cite[green,][]{Wolf2018},
\textit{Gaia} \cite[blue,][]{riello21}, and VISTA \cite[dark red,][]{McMahon2013}.   
}
    \label{fig:sed}
\end{figure}

\section{Photometric Analysis}\label{sec:phot}

\subsection{Spectral Energy Distribution}
We used the well constrained Gaia EDR3 parallax \citep[$\varpi=\parallax$,][]{gaia21} together with the archival photometry of \shortcoords\ to estimate its radius by modeling the spectral energy distribution (SED), as shown in Figure\ \ref{fig:sed}. 
This SED was constructed from average observed magnitudes over the full orbital phase, which therefore feature fairly large standard deviations. 
When \cite{Rimoldini2023} classified the star as a short period variable from \textit{Gaia} DR3 photometry, they determined standard deviations of 0.16, 0.15 and 0.17\,mag in the $G_\mathrm{BP}$, $g'$ and $G_\mathrm{RP}$ bands, respectively. 

The best-fit atmospheric parameters obtained from the spectral fit in Section \ref{sect:spec} were kept fixed for the SED fit. 
Free parameters were the angular diameter $\log \Theta/\mathrm{rad} = -11.216^{+0.035}_{-0.029}$ and the interstellar reddening, which turned out to be consistent with zero.  

The donor radius was then be computed via $R_2 = \varpi / \Theta$ to be $\rb$.
The donor mass can be derived from $M = g R^2 / G$, where $g$ is the spectroscopic surface gravity and $G$ is the gravitational constant. 
The minimum mass of the accretor is then constrained through the binary mass function from the radial velocity curve. We find that \shortcoords\ contains a bright low mass pre-white dwarf donor with $M_2=\mb$ and a faint ultra-massive white dwarf accretor with $M_1>\minmass$, based on our spectroscopic analysis.

\begin{figure*}
    \centering
    \includegraphics[width=\textwidth]{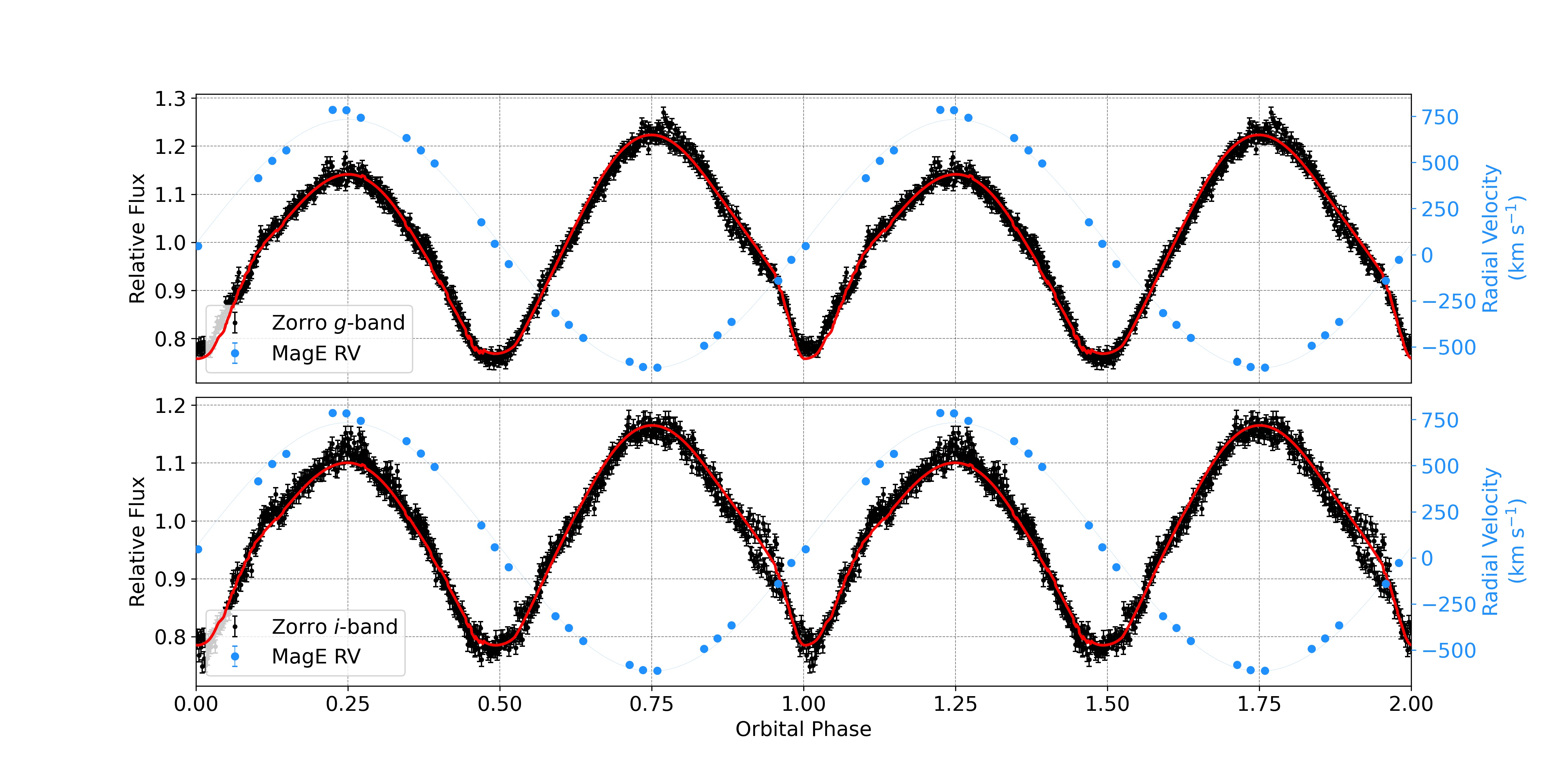}
    \caption{Black points: Gemini South Zorro $g$-band (top) and $i$-band (bottom) phase-folded light curves for \shortcoords. Red line: best-fitting model light curve obtained through our \textsc{lcurve} modeling. Blue points (right axes): observed radial velocity values from our MagE optical spectroscopy.}
    \label{fig:zorro}
\end{figure*}

\subsection{Gemini South Photometry}
\shortcoords\ (TIC 938779482) shows periodic photometric variability in the 2-minute cadence data of TESS Sector 37. However, because TESS has a large pixel scale ($21~{\rm arcsec~px^{-1}}$), the photometric variability seen TESS is diluted by many nearby field stars in the relatively crowded field, which contains 10 other objects brighter than Gaia $G=18.0~{\rm mag}$ within $1~{\rm arcmin}$.

We obtained 61 minutes of simultaneous dual-filter $g'+i'$ time-series photometry of \shortcoords\ using Gemini South's Zorro instrument, covering two full binary orbits with an exposure time of 3-seconds on UT 2024 July 15. Due to an issue with the observing sequence, our data included 2.0-second or 3.0-second gaps between each individual exposure, which would otherwise not be present in high-speed photometry using the Zorro instrument.

We reduced our Zorro data using standard CCD photometry techniques, including bias subtraction and flat field correction. We performed relative aperture photometry to extract the light curve of \shortcoords\ and calibrated it using the average light curve of at least five nearby bright field stars. Our calibrated Zorro $g'$-band and $i'$-band light curves show evidence for 1) a tidally distorted visible component (the low mass donor star), 2) deep eclipse features from an accretion disc and donor star, and 3) a bright spot from the accretion stream impacting the accretion disc. Our phase-folded Zorro light curves are displayed in Figure \ref{fig:zorro}. We added an additional $1\%$ flux uncertainty in quadrature to our flux error bars to account for small-amplitude variations in the dynamically variable accretion disc (disc flickering).

\subsection{Light curve modeling}
We performed a Markov Chain Monte Carlo (MCMC) modeling analysis of our light curves using \textsc{lcurve} \citep{copperwheat2010} to simultaneously fit our Zorro $g'$-band and $i'$-band light curves, including components for the accretion disc and the accretion bright spot. We used filter-dependent gravity darkening and quadratic limb darkening coefficients from \citet{claret2020} for a $\log{g_2}=\loggbldc$, $T_{\rm eff,2}=\teffbldc$ donor and a $\log{g_1}=8.5$, $T_{\rm eff,1}=10,000~{\rm K}$ accretor. We placed Gaussian priors on the donor velocity semi-amplitude ($K_2$), radius ($R_2$), and effective temperature ($T_{\rm eff,2}$) based on our spectroscopic and SED analysis.

We allowed the following \textsc{lcurve} parameters to vary in our fit: mass ratio ($q$), orbital inclination ($i$), donor temperature ($T_{\rm eff,2}$), velocity scale ($\frac{|K_2+K_1|}{\sin{i}}$), time of primary conjunction ($T_0$), disc temperature ($T_{\rm disc}$), disc height ($h_{\rm disc}$), disc geometry beta parameter ($\beta$), radial distance between the accretor and the accretion bright spot ($\frac{d_{\rm spot}}{a}$), bright spot scale length ($\frac{l_{\rm spot}}{a}$), bright spot temperature ($T_{\rm spot}$), and the angle, tilt, and yaw of the bright spot, which define its orientation with respect to the accretion disc. We fixed the inner disc radius to be equal to the radius of the accretor and fixed the outer disc radius to the bright spot distance, which ensures that the bright spot is located at the outer edge of the accretion disc.

Our light curve modeling places constraints on the binary mass ratio and orbital inclination of \shortcoords\ based on the photometric variability and noise in our Zorro light curves. However, the precise shape and amplitude of the photometric variability appears to be heavily affected by the accretion disc and bright spot, which both vary on relatively short timescales compared to the orbital period, resulting in slightly different eclipse shapes for each orbital cycle. Quantifying this effect on the uncertainties in our best-fitting model parameters is difficult. Thus, we simply report only the computational uncertainties from our MCMC analysis.

We present our best-fitting model parameters in Table \ref{table:basic_info} and display the best-fitting model light curves overplotted onto our Zorro light curves in Figure \ref{fig:zorro}. The best-fitting donor radius, mass ratio, and inclination values, together with our observed velocity semi-amplitude ($K_2$) allow us to estimate the individual masses of the donor and accretor as $M_2=\mblcurve$ and $M_1=\malcurve$, respectively. These masses are consistent at the $1.5\sigma$ level with our spectroscopic estimates. However, given the smaller uncertainties from our light curve model fit and quality of our light curve data, we choose to use our photometric mass estimates as the true masses of \shortcoords.

\begin{table}
\center
  \renewcommand{\arraystretch}{1.3}
  \addtolength{\tabcolsep}{2pt}
	\begin{tabular}{l r}
    \hline
    \hline
    \textsc{source\_id} (Gaia DR3) & 5369709061413007360 \\
    R.A. (2016.0) & \coordra \\
    Decl. (2016.0) & \coorddec \\
    Gaia G (mag) & $16.895\pm0.006$\\
    Parallax (mas) & $1.83\pm0.06$ \\
    $\mu_{\rm R.A.}$ (mas/yr) & $-21.36\pm0.05$ \\
    $\mu_{\rm Decl.}$ (mas/yr) & $3.65\pm0.05$ \\
    \hline
    TIC & 938779482 \\
    $P_{\rm orb,TESS}$ (${\rm min}$) & $\ptessminb$ \\ 
    \hline
    $P_{\rm orb,RV}$ (${\rm min}$) & $\prvbb$ \\
    $K_2$ (${\rm km~s^{-1}}$) & $\kbb$ \\
    $\gamma$ (${\rm km~s^{-1}}$) & $\gammabb$ \\
    $T_{\rm eff,2}$ (${\rm K}$) & $\teffbb$\\
    \hline
    $q$ & $\qlcurve$ \\
    $i$ (deg) & $\incc$ \\
    $R_{\rm 2}$ (${\rm R_\odot}$) & $\rblcurvec$ \\
    $M_2$ (${\rm M_\odot}$) & $\mblcurvec$ \\
    $M_1$ (${\rm M_\odot}$) & $\malcurvec$ \\
    \hline

    \hline
    
    \hline
	\end{tabular}
    \caption{Top: Archival parameters for \shortcoords\ from Gaia DR3 and TESS. Middle: Spectroscopic constraints to \shortcoords\ based on time series MagE spectroscopy. Bottom: Best-fitting model parameters derived from an MCMC analysis to our Gemini light curve for \shortcoords. The uncertainties reported for our light curve modeling are statistical uncertainties based on the resulting parameter distributions from our MCMC simulations. Systematic uncertainties, likely at the few percent level, have not been included.}
    \label{table:basic_info}
\end{table}
\section{Discussion}\label{sec:disc}
\subsection{Gravitational Wave Simulations}
To predict the expected gravitational wave signal based on our photometric analysis for \shortcoords, we employed the parallel-tempered Markov Chain Monte Carlo algorithm, \textsc{ucb\_mcmc} within the \textsc{ldasoft}\footnote{\url{https://tlittenberg.github.io/ldasoft/html/index.html}} package \citep{littenberg2020}. The sky position and orbital period were fixed, while a Gaussian prior was applied to the distance using the Gaia DR3 parallax and a uniform prior was applied to the orbital inclination. The simulation uses an estimated astrophysical foreground from the unresolved Galactic binaries as described in \citet{cornish2017}. We ran the simulation for 3, 6, 12 and 48 months. Following the definition in Sec. 4 in \citet{kupfer2024}, which states that a source is detected in LISA when the posterior distribution shows a closed contour, the source will be detected after 6 months of LISA observations. After 48 months of LISA observations, we predict an uncertainty for the inclination of $\Delta\iota\approx5^\circ$ and a precision for the gravitational wave amplitude ($\mathcal{A}$) of $\sigma_\mathcal{A}/\mathcal{A}\approx26\%$. The chirp mass is expected to be measured with $\mathcal{M} = 0.403\pm0.013$ which is more precise than the current measurement in this work. Using the {\sc LEGWORK}\footnote{\url{https://legwork.readthedocs.io/en/latest/}} module \citep{wagg2022a} we predict a signal-to-noise $S/N\approx7-10$ after 48 months of LISA observations.

\subsection{Future Evolution and Merger Outcome}

The evolutionary fate of an ultra-compact binary is determined by the physics affecting its orbital angular momentum. General relativity predicts that the emission of gravitational waves will cause the binary orbit to shrink, eventually leading to a merger. However, active accretion in a degenerate binary may instead lead to a widening orbit after reaching a period minimum, avoiding a merger and resulting in a stable accreting binary, such as an AM CVn type system. At much shorter orbital periods ($P_{\rm orb}\lesssim10~{\rm min}$), tidal interactions begin to play a non-negligible role in orbital evolution of ultra-compact binaries.

If only gravitational wave emission is considered, we expect that \shortcoords\ will experience rapid orbital decay, leading to a direct merger event within $\tau_{\rm merge}=\tmerge$, based on our derived system parameters. In this case, the merger is expected to produce a sub-Chandrasekhar Type Ia supernova \cite[see][]{fink2010}.

Depending on the accretion rate and the remaining hydrogen layer mass of the donor star, hydrogen nova outbursts on the accretor may accelerate the orbital decay and lead to a direct merger as the donor experiences dynamical friction while orbiting through the ejected material. \citet{shen2015} has demonstrated that potentially all interacting double white dwarf binaries may eventually merge in this way.

An X-ray detection may provide an estimate to the accretion rate of \shortcoords. However, we note that eROSITA DR2 does not detect an X-ray counterpart to \shortcoords\ to within an upper limit of $8.9\times10^{-14}~{\rm erg~cm^{-2}~s^{-1}}$ between $0.2-2.3~{\rm keV}$ \citep{tubinarenas2024,merloni2024}. Thus, we are unable to place precise constraints on the accretion rate to determine its effect on the orbital evolution. Nevertheless, our data clearly show evidence for active accretion of hydrogen-rich material from the donor star. The eventual depletion of the donor's hydrogen atmosphere will lead to accretion of helium-rich material onto the massive white dwarf. This helium-rich accretion will lead to the detonation of the helium layer on the surface of the accretor, which will likely trigger a detonation of the CO-core, also resulting in a Type Ia supernova explosion \citep{shen2024}.

In any case, the eventual fate of \shortcoords\ is almost certain to be a Type Ia supernova explosion within the next few million years. This makes \shortcoords\ the first well-constrained LISA-detectable Type Ia supernova progenitor discovered.

\section{Conclusions}\label{sec:conc}

We have presented the discovery of a new bright, nearby, ultra-compact, accreting binary in the southern sky based on selections to archival photometry from SkyMapper and astrometry from Gaia. Our analysis of new time-series spectroscopy and photometry data finds that \shortcoords\ will almost certainly explode as a Type Ia supernova through the double-detonation channel, even if not through a direct merger due to gravitational wave emission.

Future eclipse timing monitoring of \shortcoords\ will enable an independent estimate of its chirp mass ($\mathcal{M}$) and orbital period derivative ($\dot{P}$) based on an observed decay or widening of the orbit. A precise measurement of $\dot{P}$, combined with our photometrically-derived masses, would additionally provide an estimate to the accretion rate based on the gravitational wave contribution to an observed $\dot{P}$.

We note that the ATLAS DR4 forced photometry service \citep{shingles2021} provides historical photometry on \shortcoords\, but the quality and cadence of the data make it difficult to estimate precise mid-eclipse timings for use towards obtaining a precise period derivative. Instead, we leave determining the precise $\dot{P}$ of \shortcoords\ to a future work.

Our work here has shown that there are still bright nearby ultra-compact binaries yet undiscovered in the southern sky. Similar to ZTF in the north, future southern sky time domain surveys will enable rapid discovery of these ultra-compact binaries through their periodic photometric variability.

\section*{Acknowledgements}

AK acknowledges support from NASA through grant 80NSSC22K0338.

MD is supported by the Deutsches Zentrum für Luft- und Raumfahrt (DLR) through grant 50-OR-2304. 

This work is supported in part by the NSF under grant AST-2205736, the NASA under grants 80NSSC24K0436, 80NSSC22K0479, and 80NSSC24K0380.

This paper includes data gathered with the 6.5 meter Magellan Telescopes located at Las Campanas Observatory, Chile.

The authors acknowledge the High Performance Computing Center\footnote{\url{http://www.hpcc.ttu.edu}} at Texas Tech University for providing computational resources that have contributed to the research results reported within this paper.

This work has made use of data from the European Space Agency (ESA) mission {\it Gaia} (\url{https://www.cosmos.esa.int/gaia}), processed by the {\it Gaia} Data Processing and Analysis Consortium (DPAC, \url{https://www.cosmos.esa.int/web/gaia/dpac/consortium}). Funding for the DPAC has been provided by national institutions, in particular the institutions participating in the {\it Gaia} Multilateral Agreement.

Based on observations obtained at the international Gemini Observatory, a program of NSF NOIRLab, which is managed by the Association of Universities for Research in Astronomy (AURA) under a cooperative agreement with the U.S. National Science Foundation on behalf of the Gemini Observatory partnership: the U.S. National Science Foundation (United States), National Research Council (Canada), Agencia Nacional de Investigaci\'{o}n y Desarrollo (Chile), Ministerio de Ciencia, Tecnolog\'{i}a e Innovaci\'{o}n (Argentina), Minist\'{e}rio da Ci\^{e}ncia, Tecnologia, Inova\c{c}\~{o}es e Comunica\c{c}\~{o}es (Brazil), and Korea Astronomy and Space Science Institute (Republic of Korea).

Some of the observations in this paper made use of the High-Resolution Imaging instrument Zorro (program ID: GS-2024A-FT-105). Zorro was funded by the NASA Exoplanet Exploration Program and built at the NASA Ames Research Center by Steve B. Howell, Nic Scott, Elliott P. Horch, and Emmett Quigley. Zorro was mounted on the Gemini South telescope of the international Gemini Observatory, a program of NSF NOIRLab, which is managed by the Association of Universities for Research in Astronomy (AURA) under a cooperative agreement with the U.S. National Science Foundation. on behalf of the Gemini partnership: the U.S. National Science Foundation (United States), National Research Council (Canada), Agencia Nacional de Investigación y Desarrollo (Chile), Ministerio de Ciencia, Tecnología e Innovación (Argentina), Ministério da Ciência, Tecnologia, Inovações e Comunicações (Brazil), and Korea Astronomy and Space Science Institute (Republic of Korea).

This paper includes data collected with the TESS mission, obtained from the MAST data archive at the Space Telescope Science Institute (STScI). Funding for the TESS mission is provided by the NASA Explorer Program. STScI is operated by the Association of Universities for Research in Astronomy, Inc., under NASA contract NAS 5–26555. All TESS data used for this work can be found in MAST: \dataset[doi:10.17909/t9-nmc8-f686]{https://dx.doi.org/10.17909/t9-nmc8-f686}

The national facility capability for SkyMapper has been funded through ARC LIEF grant LE130100104 from the Australian Research Council, awarded to the University of Sydney, the Australian National University, Swinburne University of Technology, the University of Queensland, the University of Western Australia, the University of Melbourne, Curtin University of Technology, Monash University and the Australian Astronomical Observatory. SkyMapper is owned and operated by The Australian National University's Research School of Astronomy and Astrophysics. The survey data were processed and provided by the SkyMapper Team at ANU. The SkyMapper node of the All-Sky Virtual Observatory (ASVO) is hosted at the National Computational Infrastructure (NCI). Development and support of the SkyMapper node of the ASVO has been funded in part by Astronomy Australia Limited (AAL) and the Australian Government through the Commonwealth's Education Investment Fund (EIF) and National Collaborative Research Infrastructure Strategy (NCRIS), particularly the National eResearch Collaboration Tools and Resources (NeCTAR) and the Australian National Data Service Projects (ANDS).

This work made use of Astropy:\footnote{http://www.astropy.org} a community-developed core Python package and an ecosystem of tools and resources for astronomy \citep{astropy2013, astropy2018, astropy2022}. 

This work has made use of data from the Asteroid Terrestrial-impact Last Alert System (ATLAS) project. The Asteroid Terrestrial-impact Last Alert System (ATLAS) project is primarily funded to search for near earth asteroids through NASA grants NN12AR55G, 80NSSC18K0284, and 80NSSC18K1575; byproducts of the NEO search include images and catalogs from the survey area. This work was partially funded by Kepler/K2 grant J1944/80NSSC19K0112 and HST GO-15889, and STFC grants ST/T000198/1 and ST/S006109/1. The ATLAS science products have been made possible through the contributions of the University of Hawaii Institute for Astronomy, the Queen’s University Belfast, the Space Telescope Science Institute, the South African Astronomical Observatory, and the Millennium Institute of Astrophysics (MAS), Chile.

\facilities{Gemini South (Zorro), Magellan:Baade (MagE spectrograph),} 
\software{\textsc{astropy} \citep{astropy2013,astropy2018,astropy2022},
          \textsc{iraf} \citep{tody1986,tody1993},
          \textsc{lcurve} \citep{copperwheat2010},
          \textsc{isis} \citep{Houck2000},
          \textsc{ldasoft} \citep{littenberg2020},
          \textsc{legwork} \citep{wagg2022b}.
          }

\bibliographystyle{aasjournal}

\begin{appendix}

\section{Spectral fits by phase}
\label{appendix:spec}

Here we present the phase-resolved spectral fits for select MagE spectra of \shortcoords\ at notable orbit phase ($\phi=0, 0.47, 0.76$). The observation is shown in black, the best-fit model in red, the donor contribution in blue, and the disc contribution in dark red. Grayed out regions were not considered in the $\chi^2$ fit. More details are given in Section \ref{sect:spec}. 

\begin{figure}
\centering
\includegraphics[width=23.2cm,angle=90,page=1]{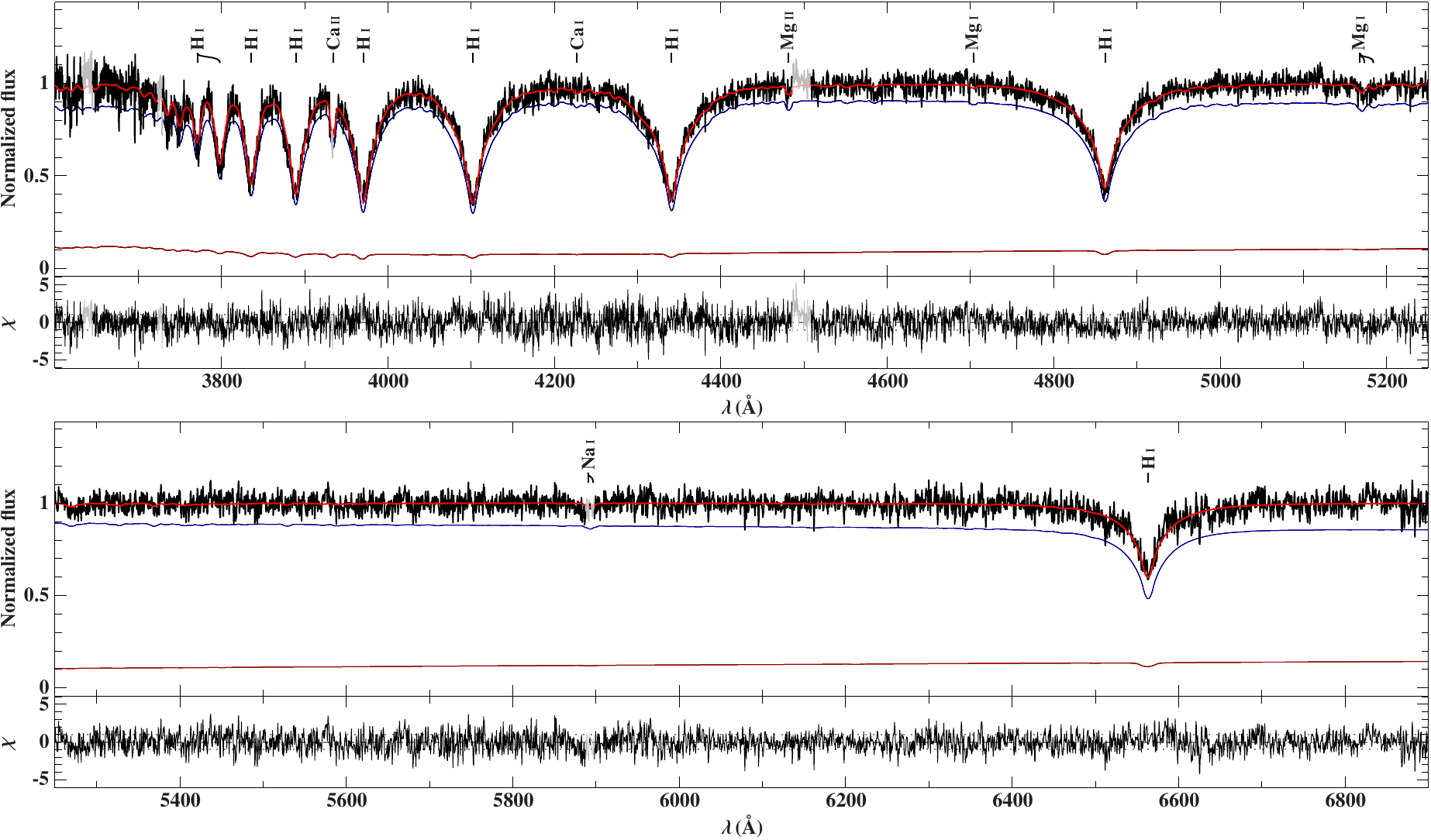}
\caption{MagE spectrum (black) and the combined (red) and individual models at $\phi = 0$.}
\label{fig:hrs:full}
\end{figure}
\begin{figure}
\centering
\includegraphics[width=23.2cm,angle=90,page=11]{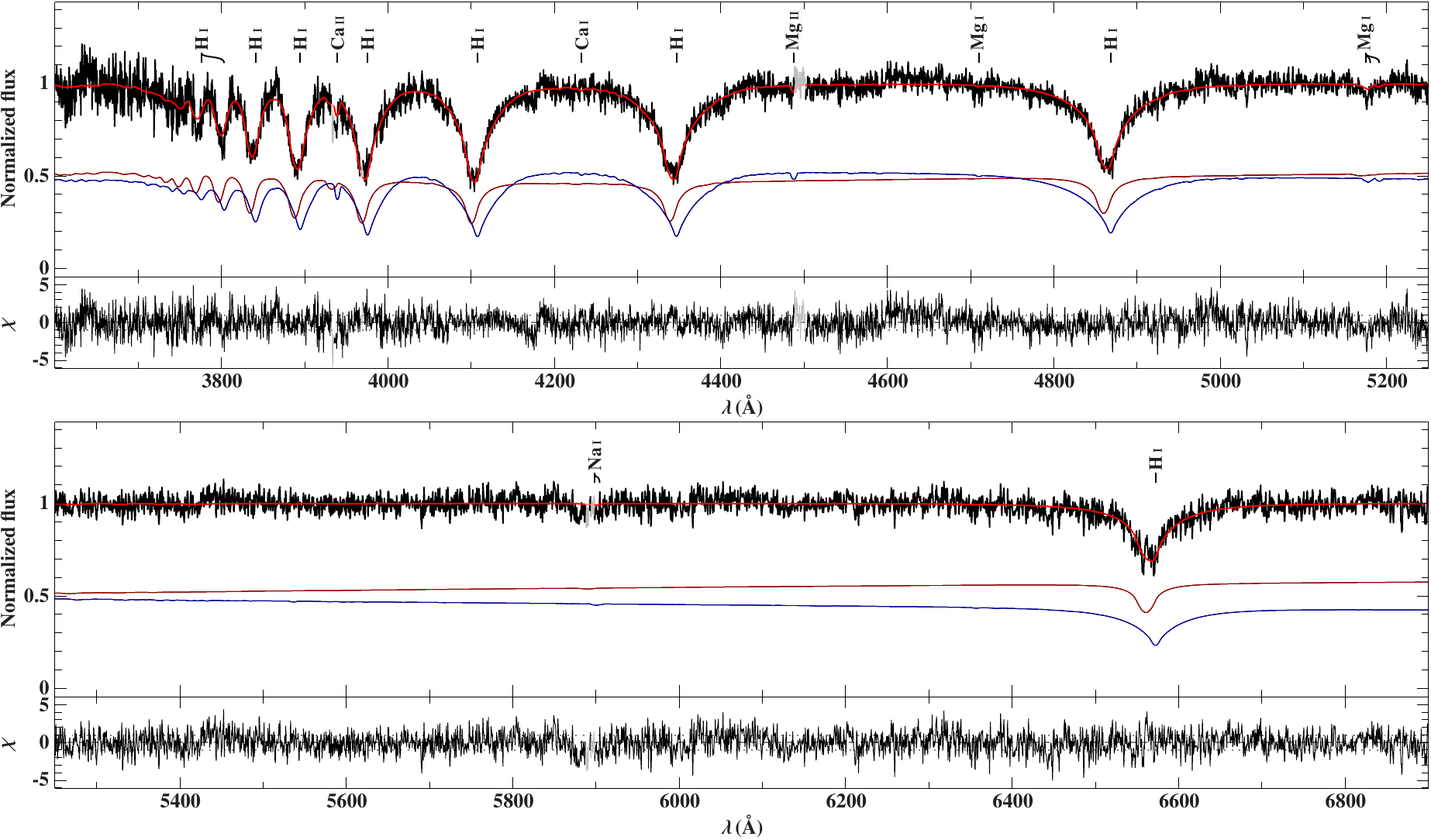}
\caption{MagE spectrum  (black) and the combined (red) and individual models  at $\phi = 0.47$.}
\label{fig:hrs:full2}
\end{figure}
\begin{figure}
\centering
\includegraphics[width=23.2cm,angle=90,page=19]{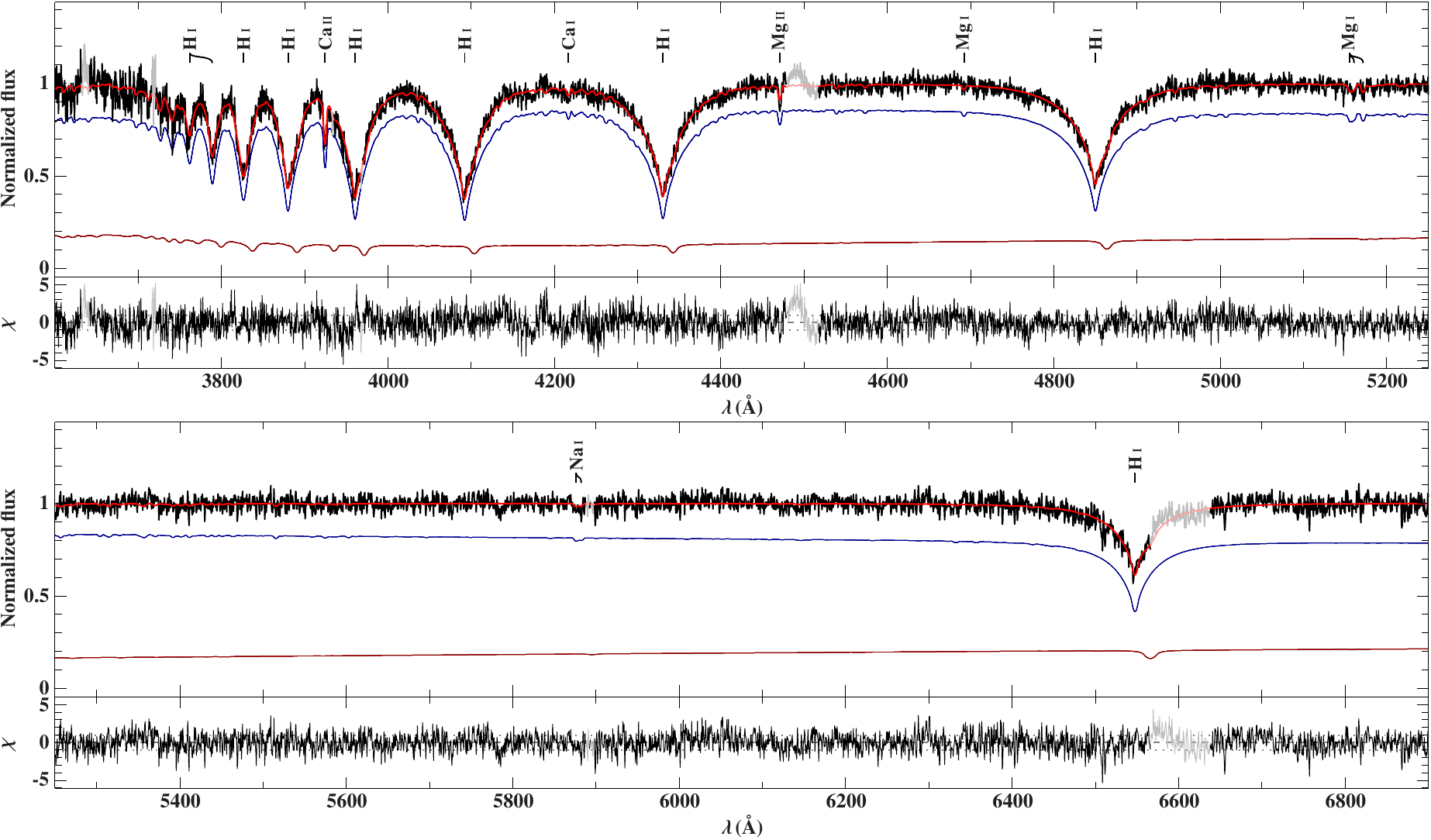}
\caption{MagE spectrum (black) and the combined (red) and individual models at $\phi = 0.76$. }
\label{fig:hrs:full3}
\end{figure}

\end{appendix}

\end{document}